# Superconductivity at 28.3 K and 17.1 K in $(Ca_4Al_2O_{6-y})(Fe_2Pn_2)$ (*Pn* = As and P)

Parasharam M. Shirage[1,†], Kunihiro Kihou[1,2], Chul-Ho Lee[1,2], Hijiri Kito[1,2], Hiroshi Eisaki[1,2] and Akira Iyo[1,2,*]

[1]National Institute of Advanced Industrial Science and Technology, Tsukuba, Ibaraki 305-8568, Japan

[2]JST, Transformative Research-Project on Iron Pnictides (TRIP), 5, Sanbancho, Chiyoda, Tokyo 102-0075, Japan

†E-mail: paras-shirage@aist.go.jp

*E-mail: iyo-akira@aist.go.jp

**Abstract:** Here we report on discovery of $(Ca_4Al_2O_{6-y})(Fe_2Pn_2)$ (*Pn* = As and P) (Al-42622(*Pn*)) superconductor using high-pressure synthesis technique. Al-42622(*Pn*) exhibit superconductivity for both *Pn* = As and P with the transition temperatures of 28.3 K and 17.1 K, respectively. The *a*-lattice parameters of Al-42622(*Pn*) (*a* = 3.713 Å and 3.692 Å for *Pn* = As and P, respectively) are smallest among the iron-pnictide superconductors. Correspondingly, Al-42622(As) has the smallest As-Fe-As bond angle (102.1 °) and the largest As distance from the Fe planes (1.500 Å).

One of the salient features of the recently-discovered iron (Fe)-based superconductors is their wide material variation. Their basic crystal structure is an alternative stacking of the Fe*Pn* (*Pn*=P, As) layers sandwiched by the blocking layers. Variety of crystal structures can be realized by employing different blocking layers, such as *Ln*O (*Ln*=rare earth)[1], *Ae* (*Ae*=Ba, Sr, Ca, Eu)[2], and *A* (*A*=Li, Na) [3]. Most recent achievement is the discovery of superconductors possessing perovskite-type blocking layers, $(Sr_4Sc_2O_6)(Fe_2P_2)$[4], $(Sr_4M_2O_{6-y})(Fe_2As_2)$ (*M* = V[5] and MgTi[6]), $(Sr_4M_3O_8)(Fe_2As_2)$ (*M* = ScTi ), $(Ba_4Sc_3O_{7.5})(Fe_2As_2)$ [7], $(Ca_{n+m}(M,Ti)_nO_y)(Fe_2As_2)$ (*M* = Sc[8], Mg[9,10] and Al[11], $n$ = 2, 3, 4 and 5, $m$ = 1, 2 ), with $T_c$ reaching as high as 47 K for $(Ca_4(Mg,Ti)_3O_y)(Fe_2As_2)$. Another interesting feature is the strong correlation between their superconducting transition temperature ($T_c$) and the configuration of the $FeAs_4$ local structure, such as the *Pn*-Fe-*Pn* bond angle ($\alpha$) and/or the height of *Pn* atoms relative to the neighboring Fe layers ($h_{Pn}$)[12, 13] and many theories have been proposed to account for the correlation [14, 15]. Considering these situations, synthesis of materials possessing either (1) new type of blocking layers or (2) unachieved $FeAs_4$ configurations, should be tremendously beneficial for understanding the high-$T_c$ mechanism, and, eventually, for enhancing $T_c$, of the Fe-based superconductors.

Based on the above background, in this study we have attempted to synthesize perovskite-type Fe*Pn* superconductors using a high-pressure (HP) technique. HP technique is established as a versatile tool to explore new materials beyond the limitation of ambient synthesis conditions [16]. By enforcing the stacking between the FeAs layers and the blocking layers with smaller in-plane unit cell length (*a*-lattice parameters), this method allows us to synthesize materials with extremely smaller unit cell volumes. For example, we have succeeded in synthesizing a series of $LnFeAsO_{1-y}$ superconductors ($Ln$=La, Ce, Pr, Nd, Sm, Gd, Tb, Dy) that possess small $Ln$ atoms, which yield small $Ln$O blocking layers [17]. The unit cell volume largely shrinks from 141 Å$^3$ ($Ln$ = La) to as small as 124 Å$^3$ ($Ln$ = Dy). In this study, we have applied the HP technique in synthesizing materials containing perovskite-type blocking layers which are extremely small and therefore cannot be synthesized under ambient conditions. Along this line, we have discovered a series of (Ca$_4$Al$_2$O$_{6-y}$)(Fe$_2$Pn$_2$) ($Pn$ = As and P) (abbreviated as Al-42622($Pn$)) superconductors with $T_c$ = 28.3 K and 17.1 K, respectively. As a consequence of HP synthesis, the synthesized compounds possess the unique structural characteristics never realized in the existing Fe-based superconductors, namely, shortest $a$-axis lattice parameters, smallest As-Fe-As bond angle, and longest As-Fe plane distance. Here we demonstrate the synthesis method and the characterization of Al-42622($Pn$).

Polycrystalline samples of Al-42622($Pn$) were synthesized by the solid-state reaction method using a cubic-anvil-type HP apparatus. It should be noted that one cannot synthesize Al-42622($Pn$) by conventional ambient pressure synthesis [11]. Starting materials were powders of CaO, Al, As, P, Fe and Fe$_2$O$_3$. The powders were mixed with a nominal composition of (Ca$_4$Al$_2$O$_{6-y}$)(Fe$_2$Pn$_2$). Here the oxygen content 6-$y$ was controlled by adjusting the ratio of Fe and Fe$_2$O$_3$. We have found that the purity of the obtained samples strongly depends on the oxygen content in the starting composition. The suitable oxygen content turned out to be 6 - $y$ = 5.6 ~ 5.8. The starting materials were ground using an agate mortar and pressed into a pellet in a glove box filled with N$_2$ gas. The pellet was loaded in a high-pressure cell and heated at 1150 and 1300 °C for $Pn$ = As and P, respectively, under a pressure of 4.5 GPa for 1 h.

Powder X-ray diffraction (XRD) patterns were measured using CuK$_\alpha$ radiation.

Neutron scattering measurement was carried out using the high-resolution powder diffractometer HERMES of the Institute for Materials Research, Tohoku University, installed at the JRR-3 reactor of JAEA in Japan. Incident neutron wavelength was fixed at 1.8484 Å using a Ge monochromator. The data were analyzed by the Rietveld method using the Rietan program. The dc magnetic susceptibility was measured using a SQUID magnetometer (Quantum Design MPMS) under a magnetic field of 5 Oe. The resistivity was measured by a four-probe method.

Fig. 1 represents the XRD patterns of Al-42622($Pn$) samples with nominal compositions of (a) $Ca_4Al_2O_{5.65}Fe_2As_2$ and (b) $Ca_4Al_2O_{5.80}Fe_2P_2$, respectively. Major peaks can be indexed based on the tetragonal crystal structure with $a$ = 3.713 Å and $c$ = 15.407 Å for Al-42622(As) while with $a$ = 3.692 Å and $c$ = 14.934 Å for Al-42622(P). Small impurity phases of CaO, and FeAs were detected in the Al-42622(As) samples. Note that the $a$-lattice parameter of Al-42622(As) is significantly smaller than those of Fe-As related compounds containing Ca-based perovskite-derived blocking layers, such as $(Ca_4(Mg,Ti)_3O_y)(Fe_2As_2)$ ($a$ = 3.877 Å, $T_c$ = 47 K) [9] and $(Ca_5(Mg,Ti)_4O_y)(Fe_2As_2)$ ($a$ = 3.864 Å, $T_c$ = 43 K) [10]. The contraction of the $a$-lattice parameter is reasonable, considering the smaller ionic radius size of $Al^{3+}$ ($R_{Al}$ = 0.535 Å) compared to $Mg^{2+}$ ($R_{Mg}$ = 0.720 Å) and $Ti^{4+}$ ($R_{Ti}$ = 0.605 Å) [18]. Similarly, the $a$-lattice parameter of Al-42622(P) is much smaller compared to the iso-structure compound, $(Sr_4Sc_2O_6)(Fe_2P_2)$ ($a$ = 4.016 Å, $T_c$ = 17 K), which contains larger $Sc^{3+}$ ($R_{Sc}$ = 0.745 Å) . It is worth stressing that the $a$-parameters of Al-42622($Pn$) are smallest among ever-reported Fe-based superconductors. For example, typical materials having short $a$-lattice parameter are FeSe ($a$ = 3.769 Å, $T_c$ = 8 K)[19], LiFeAs ($a$ =3.7914 Å, $T_c$ = 18 K) [3], $(Ca,Na)Fe_2As_2$ ($a$ = 3.880 Å, $T_c$ = 26 K)[20] or $DyFeAsO_{1-y}$ ($a$ = 3.820 Å, $T_c$ = 43 K)[17].

The *c*-lattice parameters of Al-42622(*Pn*) are also significantly smaller than those of iso-structural 42622(*Pn*), such as (Sr$_4$$M_2$O$_{6-y}$)(Fe$_2$As$_2$) ($c$ = 15.809 Å for $M$ = Sc, 15.683 Å for $M$ = Cr and 15.673 Å for $M$ = V), [5, 21] due to the smaller ionic radii of Ca$^{2+}$ ($R_{Ca}$ = 1.00 Å) and Al$^{3+}$ compared to those of Sr$^{2+}$ ($R_{Sr}$ = 1.180 Å), Sc$^{3+}$ ($R_{Sc}$ = 0.745 Å), Cr$^{3+}$ ($R_{Cr}$ = 0.615 Å) and V$^{3+}$ ($R_V$ = 0.640 Å). Small lattice parameters of Al-42622(*Pn*) naturally explain why these compounds are synthesized only by HP technique. It also demonstrates the usefulness of HP technique for synthesizing materials possessing extremely smaller unit cell volumes.

Fig. 2(a-b) shows the temperature (*T*) dependence of the zero-field-cooled (ZFC) and the field-cooled (FC) magnetic susceptibility ($\chi$) for the (*a*) Al-42622(As) and (*b*) Al-42622(P) samples. As shown in the inset Fig. 2, the sample shows superconducting transition with their onset temperature ($T_c^{\chi\text{-}onset}$) at 26.8 K and 17.0 K for Al-42622(As) and Al-42622(P), respectively. The superconducting volume fraction estimated from $\chi$ (ZFC) at 5 K without demagnetizing field correction are 115 % for Al-42622(As) and 96 % for Al-42622(P), respectively, ensuring the bulk superconductivity. For Al-42622(As), the reversible region, where ZFC and FC magnetization curves overlap with each other, exisits from $T_c^{\chi\text{-}onset}$ to 21.5 K, possibly due to the grain pinning mechanism. The reversible behavior is a unique feature of the FeAs-based superconductors which contain the thick perovskite-type blocking layers, suggesting high anisotropy of their superconducting properties. In contrast, the reversible region of Al-42622(P) is extremely limited. This contrast indicates the difference of the anisotropy between the two systems, possibly because the spacing between Fe layers is shorter for Al-42622(P) compared to Al-42622(As).

In Fig. 3 *T*-dependent resistivity($\rho$) of (a) Al-42622(As) and (b) Al-42622(P) are shown. Metallic behavior ($d\rho/dT$>0) can be seen in the all *T*-range. In particular, for 42622(As), no kink in the resistivity, a signature of antiferromagnetic/orthorhombic transition, is observed. Accordingly, we can conclude that the system is away from the parent phase which possesses long-range structural/magnetic order. As indicated in the inset of Fig. 3(a), a broadening of transition (or a two-step transition) is recognized. $T$-range where the resistive broadening occurs seems to coincide with the reversible region observed in the $\chi$-$T$ curve. It may suggest that the broadening results from the flux creep in the reversible region. In Al-42622(P), very high (one order higher than Al-42622(As)) normal state resistivity and broad superconducting transition were observed. This is partly due to the poor sintering of the samples. As indicated in the insets, $T_c^{\rho-onset}$, $T_c^{\rho-mid}$ and $T_c^{\rho=0}$ are 28.3, 27.0, 23.0 K for Al-42622(As) and 17.1, 15.6, 6.0 K for Al-42622(P), respectively.

Our data indicate that the Al-42622($Pn$) systems have the shortest $a$-lattice parameters among all the Fe-based superconductors. One would naively expect that the shrinkage of the $a$-lattice parameters causes the elongation of the Fe$Pn_4$ tetrahedra along the $c$-axis. It is indeed confirmed by the neutron powder diffraction on Al-42622(As). Table I show the atomic positions of Al-42622(As) determined by the Rietveld analysis of the neutron powder diffraction. The calculated Fe-As-Fe bond angle $\alpha$ is 102.1(1)°. This value is much smaller than those of any other Fe-pnictide superconductors. For example, $\alpha$ of the $Ln$FeAsO-based materials ranges from 113° ($Ln$ = La, $T_c$ = 27 K) to 108° ($Ln$ = Dy, $T_c$ = 52 K). Similarly, $\alpha$ of (Ba,K)Fe$_2$As$_2$ ($T_c$ = 38 K) is 109°. $\alpha$ of LiFeAs ($T_c$ = 18 K) is rather smaller, but still 104°. As for those containing Ca-based perovskite-derived blocking layers, $\alpha$ ranges from 107° for Ca$_5$(Mg,Ti)$_4$O$_{11}$Fe$_2$As$_2$ [10] to 109° for Ca$_4$(Sc,Ti)$_3$O$_8$Fe$_2$As$_2$ [8], estimated from the relation between the $a$-lattice parameter and $h_{Pn}$ in ref.[22]. Empirically it is pointed out that the highest $T_c$ materials possess $\alpha$ close to 109.5°, an angle where the FeAs$_4$ tetrahedron forms the regular shape. The relatively lower $T_c$ of 28.3 K for Al-42622(As) compared to other Ca-based perovskite counterparts with $T_c$'s reaching 47 K is likely due to the elongation of the FeAs$_4$ tetrahedron. It should be also remarked that the height of As atoms relative to the Fe layers ($h_{Pn}$) is 1.500 Å, which is the highest value among the Fe-pnictide superconductors.

In contrast, $T_c$ of 17 K for Al-42622(P) is much higher than most of the existing FeP-based superconductors, such as LaFePO ($T_c$ = 4 K)[23], LiFeP ($T_c$ = 6 K)[24], *etc*. The only material which exhibits comparable $T_c$ is (Sr$_4$Sc$_2$O$_6$)(Fe$_2$As$_2$) (Sc-42622(P)), which shares the same crystal structure as Al-42622(P). Note that their *a*-lattice parameters are significantly different from each other, *a* = 3.692 Å for Al-42622(P) and *a* = 4.016 Å for Sc-42622(P), respectively. The fact that $T_c$ does not change between these two systems suggests that $T_c$ of the FeP-based superconductors is rather insensitive to the structural parameters, contrary to the FeAs-based counterpart. Another possibility is that $T_c$ of the 42622(P) system takes its maximum value when the *a*-lattice parameter falls between 3.692 Å and 4.016 Å. In this regard, it is intriguing to synthesize an alloy of Al-42622(P) and Sc-42622(P) and check whether one can enhance its $T_c$.

In summary, we have utilized HP technique to synthesize (Ca$_4$Al$_2$O$_{6-y}$)(Fe$_2$Pn$_2$) (*Pn* = As and P), which exhibit superconductivity at 28.3 K and 17.1 K for *Pn*=As and P, respectively. These materials possess the shortest *a*-lattice parameters, correspondingly the smallest $\alpha$ and the highest $h_{Pn}$ among the Fe-pnictide superconductors.

**Acknowledgements**: We thank Prof. H. Ogino for fruitful discussion. This work was supported by a Grant-in-Aid for Specially Promoted Research (20001004) from The Ministry of Education, Culture, Sports, Science and Technology (MEXT).

**Figure captions**

Figs. 1 X-ray diffraction patterns of (*a*) $Ca_4Al_2O_{6-y}Fe_2As_2$ and (*b*) $Ca_4Al_2O_{6-y}Fe_2P_2$ samples.

Figs. 2 Temperature (*T*) dependence of susceptibility ($\chi$) of (*a*) $Ca_4Al_2O_{6-y}Fe_2As_2$ and (*b*) $Ca_4Al_2O_{6-y}Fe_2P_2$ samples. Data near the superconducting transitions. $T_c^{\chi\text{-}onset}$ is determined from the intersection of the two extrapolated lines; one is drawn through $\chi$ in the normal state just above $T_c$, and the other is drawn through the steepest part of $\chi$ in the superconducting state.

Figs. 3 *T*-dependence of resistivity ($\rho$) of (*a*) $Ca_4Al_2O_{6-y}Fe_2As_2$ and (*b*) $Ca_4Al_2O_{6-y}Fe_2P_2$ samples. Data near the superconducting transitions and a determination of $T_c$ are shown in the insets. $T_c^{\rho-onset}$ is determined from the intersection of the two lines; one is drawn through $\rho$ in the normal state just above $T_c$, and the other is drawn through the steepest part of $\rho$ in the superconducting state. $T_c^{\rho-mid}$ is 50 % drop of $T_c^{\rho-onset}$ and $T_c^{\rho=0}$ is zero resistivity.

**Table I**. Atomic parameters of Al-42622(As) (space group *P*4/*nmm*) determined by a Rietveld refinement of neutron powder diffraction data at room temperature. *B* is the isotropic atomic displacement parameter. The lattice parameters are *a* = 3.7133(1) Å and *c* = 15.4035(6) Å. The *R*-factor is $R_{wp}$ = 5.261.

| Atom  | site | *x*  | *y*  | *z*        | B (Å)     |
|-------|------|------|------|------------|-----------|
| Fe    | 2a   | 3/4  | 1/4  | 0.0        | 0.47(5)   |
| As    | 2c   | 1/4  | 1/4  | 0.0974(2)  | 0.66(8)   |
| Ca(1) | 2c   | 1/4  | 1/4  | 0.7997(3)  | 0.18(9)   |
| Ca(2) | 2c   | 1/4  | 1/4  | 0.5819(3)  | 0.33(11)  |
| Al    | 2c   | 1/4  | 1/4  | 0.3181(4)  | 0.50(14)  |
| O(1)  | 4f   | 3/4  | 1/4  | 0.2969(2)  | 0.50(6)   |
| O(2)  | 2c   | 1/4  | 1/4  | 0.4340(3)  | 0.69(9)   |

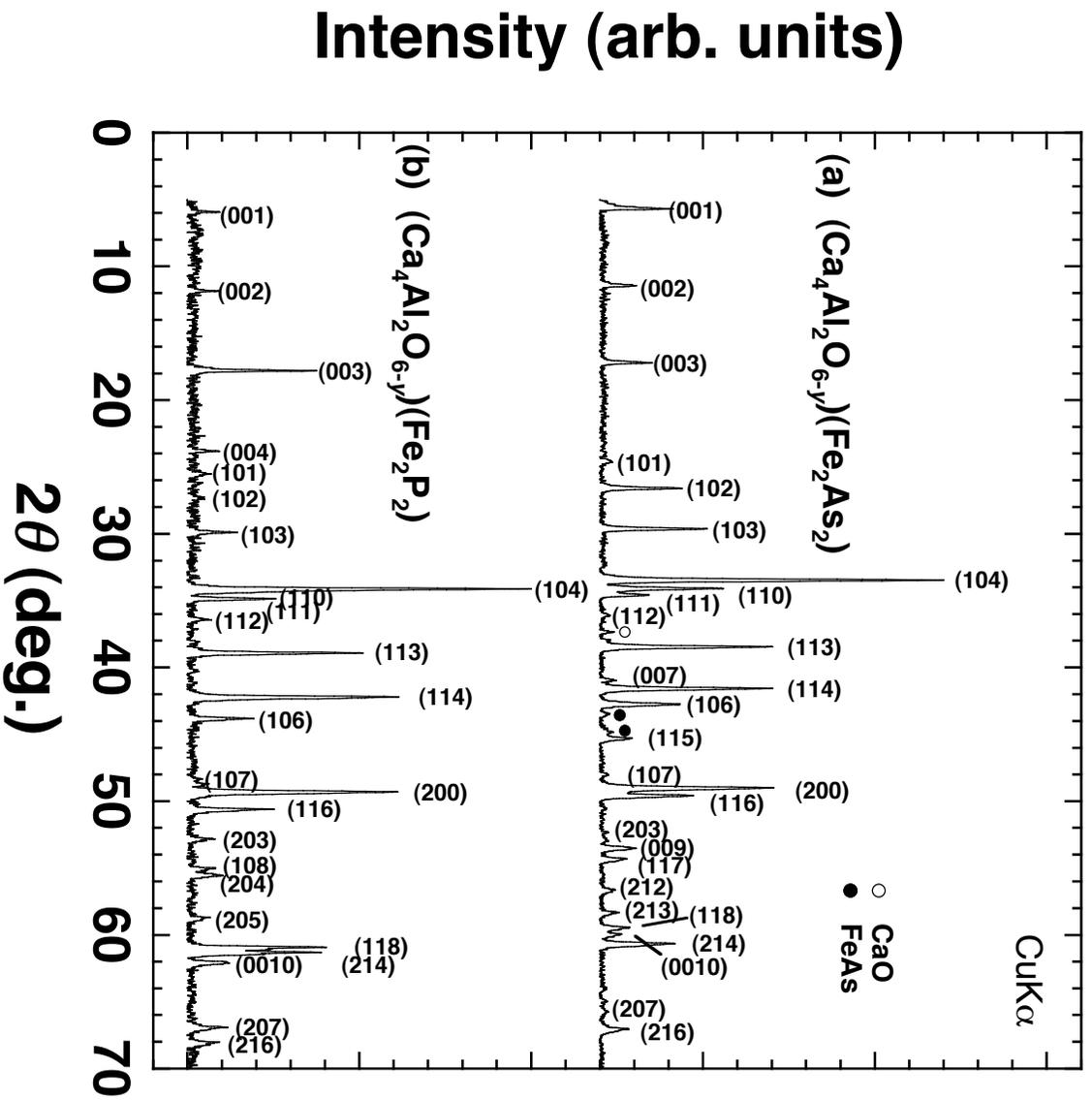

Figure 1 (a-b)

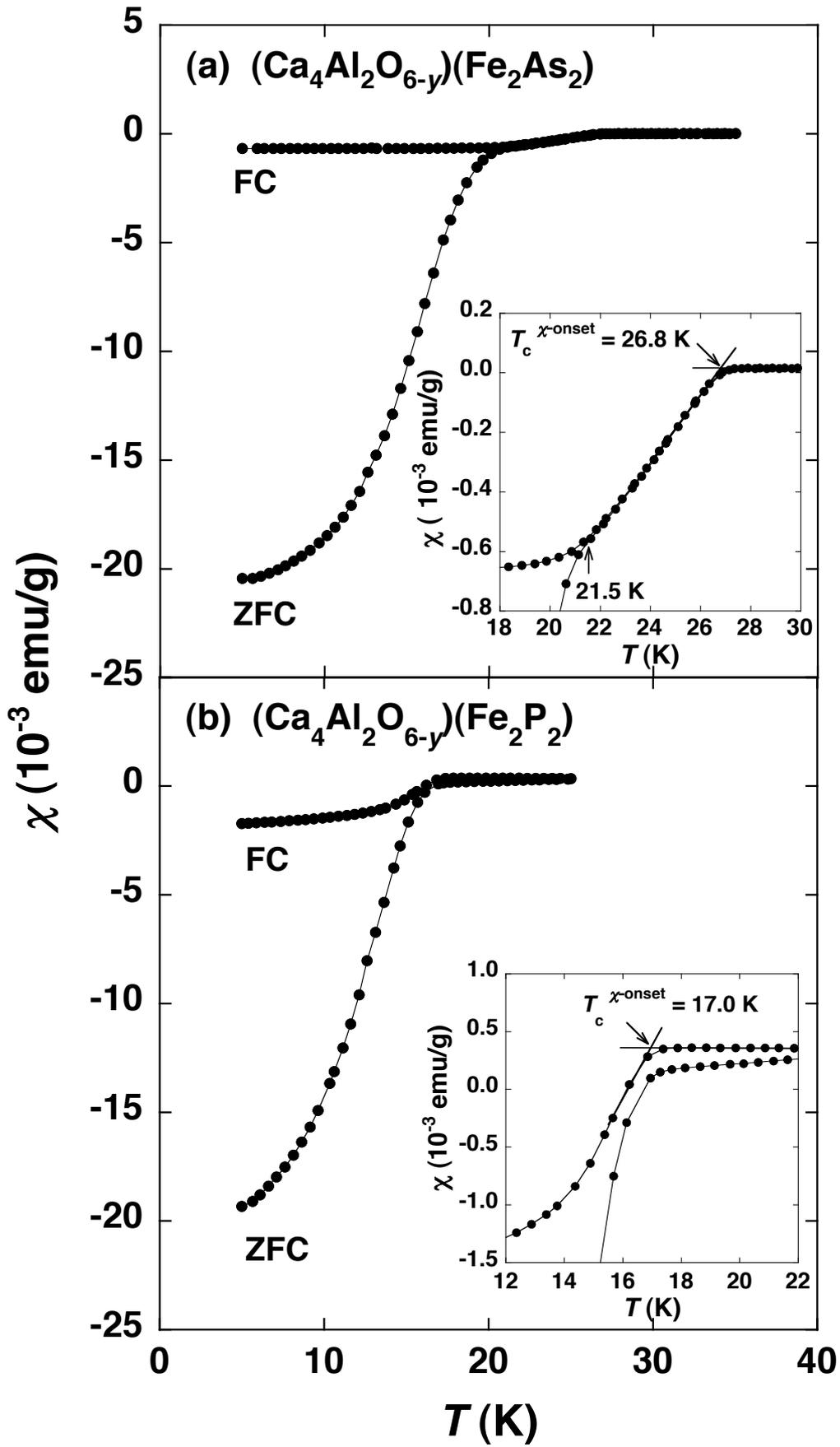

Figure 2 (a-b)

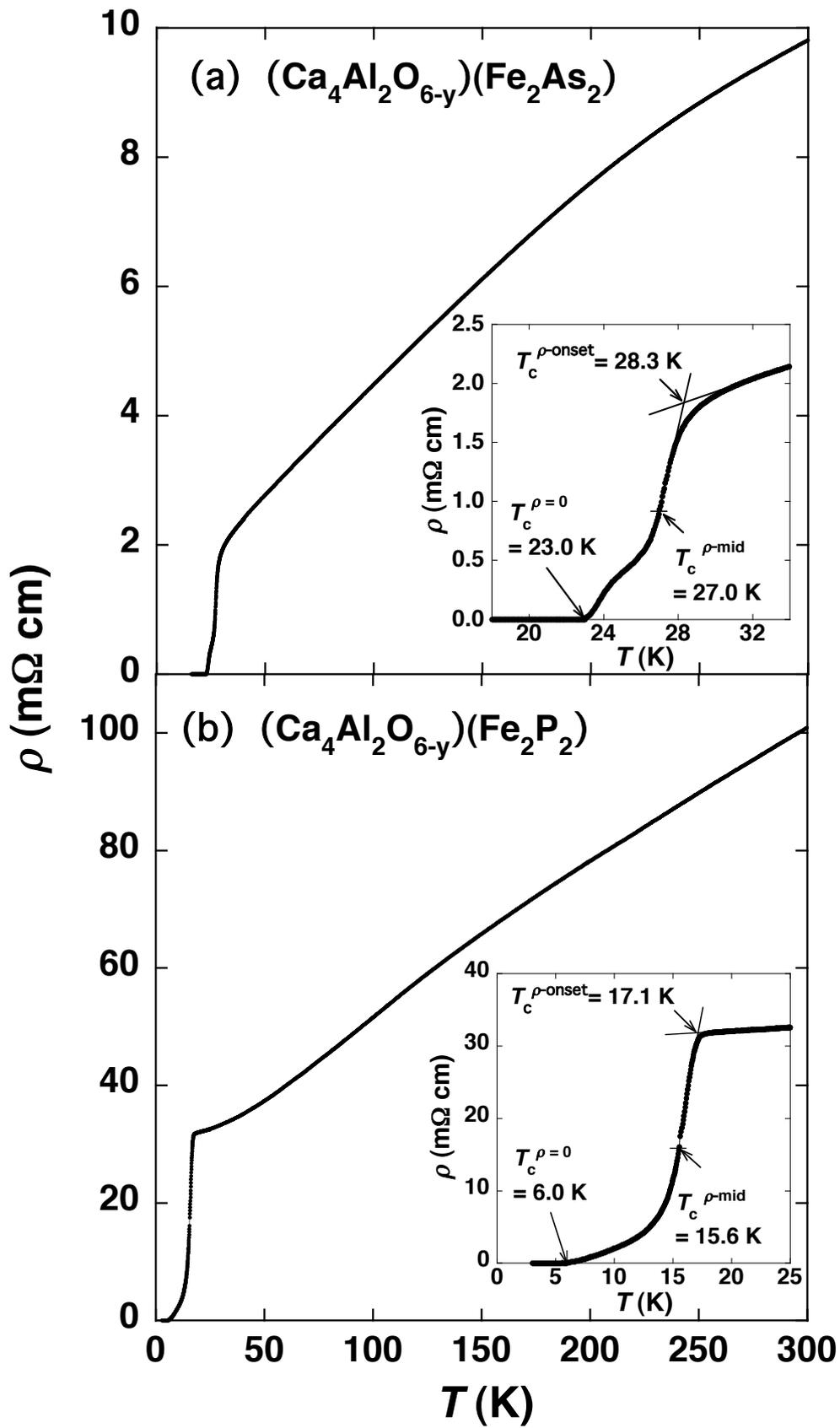

Figure 3 (a-b)